\documentclass[11pt,a4paper]{article}

\usepackage{amsmath}
\usepackage{amsfonts}
\usepackage{amssymb}
\usepackage{float}
\usepackage{graphicx}
\usepackage{booktabs}
\usepackage{authblk}
\usepackage{caption}
\usepackage{subcaption}
\usepackage[left=2.00cm, right=2.00cm, top=2.00cm, bottom=2.00cm]{geometry}
\title{Spatial deterministic wave forecasting for nonlinear sea--states}

\author[1]{M.\ Galvagno}
\author[2,3]{D.\ Eeltink}
\author[1]{R.\ Stuhlmeier \thanks{raphael.stuhlmeier@plymouth.ac.uk}}

\affil[1]{Centre for Mathematical Sciences, University of Plymouth, PL4 8AA Plymouth, UK}
\affil[2]{Department of Mechanical Engineering, Massachusetts Institute of Technology, Cambridge MA 02139-4307, USA}
\affil[3]{Department of Engineering Science, University of Oxford, Oxford OX1 3PJ, UK}
\date{}

\begin{document}
\maketitle

\begin{abstract}
We derive a simple algebraic form of the nonlinear wavenumber correction of surface gravity waves in deep water, based on temporal measurements of the water surface and the spatial Zakharov equation. This allows us to formulate an improvement over linear deterministic wave forecasting with no additional computational cost. Our new formulation is used to forecast both synthetically generated as well as experimentally measured seas, and shows marked improvements over the linear theory.
\end{abstract}

\section{Introduction}

The goal of deterministic wave forecasting is to determine what waves will arrive at a distant point, or at some future time, based on spatial or temporal measurements of the sea surface. To any beachgoer observing the erratic nature of surface water waves this may seem an impossible task, recalling Lord Rayleigh's famous statement that ``the basic law of the seaway is the lack of any law.''

However, with advances in the remote sensing of the sea surface, and attendant increases in computational power, the problem of deterministic forecasting of water waves has become tractable, and has attracted significant recent interest. 

In the present work we set out to develop the theoretical basis of a forecasting methodology that incorporates weakly nonlinear corrections to the dispersion relation up to third order. While deep water waves undergo (nearly) resonant interactions at third and higher orders, it is our goal to capture only the corrected dispersion while neglecting the slow energy exchange between wave modes. In order to compare with experiments, we consider spatial evolution, and build our theory upon the spatial Zakharov equation developed by Shemer et al \cite{Shemer2001} and employed in numerous subsequent studies \cite{Shemer2017,Kit2002,Shemer2002}.

Measurements of the free surface readily yield Fourier amplitudes, which form the basis of the linear description of surface waves. Indeed, these Fourier amplitudes can be used to construct simple and efficient linear forecasts, which have been used in practical tests of deterministic forecasting systems by Hilmer \& Thornhill \cite{Hilmer2016}, Kusters et al \cite{Kusters2016}, Al-Ani et al \cite{Al-Ani2020}, and others. The same Fourier amplitudes are the foundation of weakly nonlinear approaches, where corrections to the linear description are sought as perturbations in the (small) wave slope.

Weakly nonlinear approaches to wave forecasting include those based on PDEs like the nonlinear Schr\"odinger equation and its modifications, derived under assumptions of narrow bandwidth and used by numerous authors, including Trulsen \cite{Trulsen2005}, Simanesew et al \cite{Simanesew2017}, Klein et al \cite{Klein2020} and others. Alternatively, the well developed higher-order spectral method (HOS) \cite{Dommermuth1987,West1987} presents an attractive computational technique which has gained much recent attention by the deterministic forecasting community, see e.g.\ \cite{Wu2004a,Qi2018,Blondel2010,Guerin2019,Law2020}. 

While they better capture the evolution of real waves, the principal drawback of these approaches lies in an increased computational cost compared to linear forecasting. For practical applications, forecasts are needed on scales of seconds or minutes and tens or hundreds of meters, so speed is of the essence. Our approach is to extract the correct third-order nonlinear dispersion and include it as an essentially algebraic correction in the linear forecast. This is computationally trivial, but we will show that it yields significant advantages over the purely linear approach. 

In what follows, we first review fundamental theory, including linear forecasting, in section 2.1. Section 2.2 introduces the spatial Zakharov equation, and contains the main theoretical results. Section 3 applies linear and nonlinear forecasting methods to synthetically generated seas, simulated in a numerical wave flume using HOS. Section 4 presents comparisons with experimental measurements. Finally, section 5 presents a discussion of the results and some concluding remarks.

\section{Fundamental theory}

Assuming unidirectional propagation of long-crested waves, we may write the free surface elevation as $\eta(x,t)$ where $x$ is space and $t$ time. In order to prepare a spatial forecast the sea surface must be measured at a fixed location $x=x_0$ at $N$ times $t_0,\, t_1, \ldots t_{N-1}.$ It is simplest to assume time intervals $\Delta t = T/N$ so that $t_n = nT/N,$ but non-uniformly sampled data can be resampled using interpolation. This leads to a record
\[ y_0 = \eta(x_0,t_0), \, y_1 = \eta(x_0,t_1), \, \ldots , y_{N-1} = \eta(x_0,t_{N-1}).\]

Taking the discrete-time Fourier transform of the sequence $y_0, \, y_1, \ldots , y_{N-1}$ we find 
\begin{equation}
Y_j = \sum_{n=0}^{N-1} y_n \exp ( -i 2 \pi j n / N) = \sum_{n=0}^{N-1} y_n \exp(-i t_n \omega_j),
\end{equation}
where $\omega_j = 2 \pi j / T.$ Note that $Y(j)=Y(j+N)$ so the Fourier coefficients are $N$--periodic. The inverse transform is given by 
\begin{equation}
y_n = \frac{1}{N} \sum_{m=0}^{N-1} Y_m \exp(2 \pi i m n /N),
\end{equation}
which can be transformed into a continuous description using $nT/N \longrightarrow t$:
\begin{equation}
y(t) = \frac{1}{N} \sum_{m=0}^{N-1} Y_m \exp(i \omega_m t).
\end{equation}
Finally, this can be reformulated for $N$ even as 
\begin{equation} \label{eq: y(t)}
y(t) = \frac{Y_0}{N} + \frac{1}{N} \sum_{m=1}^{N/2-1} \left[  Y_m \exp(i \omega_m t) + Y^*_m \exp(-i \omega_m t) \right].
\end{equation}
Note that $y(t)=y(t+T).$ The term $Y_0/N = \frac{1}{N}\sum_{n=0}^{N-1} y_n$ is the mean elevation of the sampled points.

\subsection{Linear forecasting} \label{ssec: Linear forecasting}
In the linear theory of water waves there is a one-to-one correspondence between positive wavenumbers $k \in \mathbb{R}^+$ and positive frequency $\omega \in \mathbb{R}^+$, given by the dispersion relation
\[ \omega^2 = g k \tanh(k d), \]
where $g$ is the acceleration of gravity and $d$ is the (constant) water depth. For deep water $d\rightarrow \infty$ this dispersion relation reduces to the simpler expression $\omega^2 = gk.$ This correspondence allows for a linear forecast to be constructed from the $N$ samples captured in \eqref{eq: y(t)}. By stipulating that a wave with measured frequency $\omega_m$ has wavenumber $k_m = \omega_m^2/g$, it is immediately possible to write:
\begin{equation} \label{eq: zeta linear}
\zeta_L(x,t) = \frac{Y_0}{N} + \frac{1}{N} \sum_{m=1}^{N/2-1} \left[  Y_m \exp(i(k_m x - \omega_m t)) + Y^*_m \exp(-i(k_m x -  \omega_m t)) \right].
\end{equation}
The waves accounted for in the forecast then have frequencies between $\omega_1 = \frac{2\pi}{T}$ and $\omega_{N/2-1} = \frac{2\pi(N/2-1)}{T}.$ The energy associated with a given frequency moves at the group velocity, defined as 
\[ c_g := \frac{d \omega(k)}{d k},\]
with the simple form in deep water
$ c_g = 0.5{g}/{\omega}. $

\begin{figure}
\centering
\includegraphics[scale=0.7]{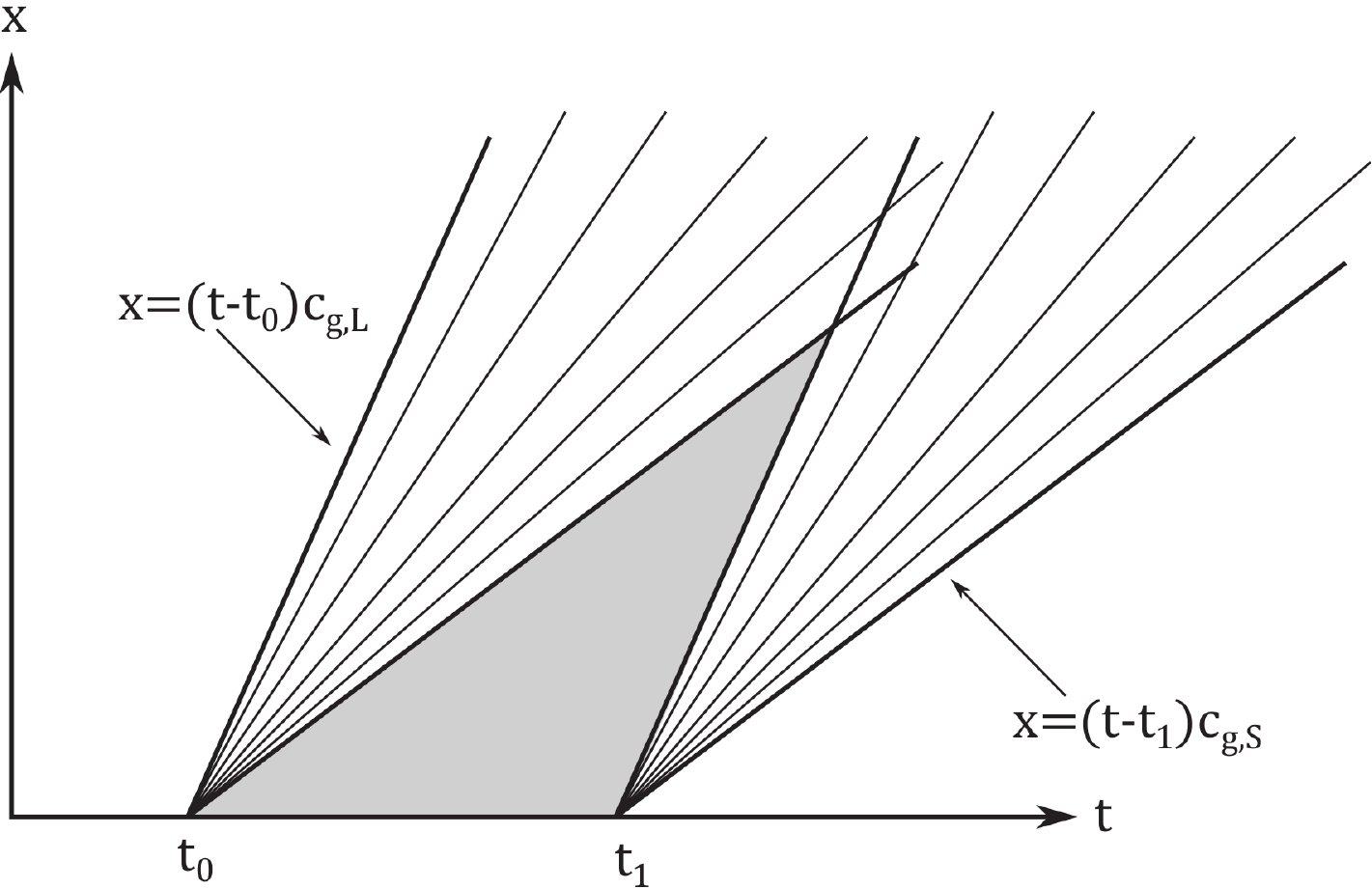}
\caption{Predictable region (grey shaded area) based on measurements at $x=0$ in $[t_0,t_1]. $}
\label{fig:Predictable region}
\end{figure}

For a given measurement, the longest waves of interest $\omega_L$ will travel fastest, and the shortest waves $\omega_S$ slowest (note that practically $\omega_L$ may not be $\omega_1,$ nor $\omega_S$ be $\omega_{N/2-1},$ as there may be negligible energy associated with the longest or shortest waves that can be theoretically resolved). This leads to the concept of a predictable region in $(x,t)$ as shown in figure \ref{fig:Predictable region}.

The thick lines in figure \ref{fig:Predictable region} show the group velocities $c_{g,L}$ and $c_{g,S}$ of the longest and shortest waves $\omega_{L}$ and $\omega_{S},$ respectively. Thinner lines in between these indicate the group velocities of waves of length intermediate between $\omega_{L}$ and $\omega_{S}.$ For a measurement at $x=0$ over time $t=[t_0,t_1],$ all the waves in the shaded region originate in the measurement domain, and are therefore \textit{predictable.} The only exceptions are waves longer than $\omega_L$ or shorter than $\omega_S$ that may encroach from $t<t_0$ or $t>t_1,$ and are not accounted for in the forecast.

\subsection{Nonlinear forecasting} \label{ssec: Nonlinear forecasting}

\subsubsection{The spatial Zakharov equation}
\label{sssec: spatial ZE}

The discussion in the preceding section \ref{ssec: Linear forecasting} is relevant only for waves of small steepness, such that linear wave theory may be employed. Weakly nonlinear theory (to third order) makes for  dramatic changes to the dispersion relation of waves in deep water, and complicates the forecast problem considerably. 

The theoretical basis for our nonlinear forecast will be the spatial Zakharov equation developed in the early 2000s by Shemer et al \cite{Shemer2001}. This takes the form
\begin{align}\nonumber
i c_g \frac{\partial B(x,\omega)}{\partial x} = &\iiint T(k,k_1,k_{2},k_{3}) B^*(x,\omega_1) B(x,\omega_2) B(x,\omega_3) \\
&\cdot  \exp(-i(k+k_1-k_2-k_3)x) \delta(\omega_1+\omega_2 - \omega_3 - \omega_4) d \omega_1 d \omega_2 d \omega_3.
\label{eq:Spatial ZE}
\end{align}
where $c_g$ denotes the deep-water, linear group velocity. This equation can be discretised as follows:
\begin{equation}
\label{eq:Spatial ZE Discr}
i c_{g,j} \frac{d B_j(x)}{dx} = \sum_{l,m,n} T_{jlmn} B_l^* B_m B_n \exp(-i(k_i + k_j - k_m - k_n)x) \delta(\omega_j + \omega_l - \omega_m - \omega_n),
\end{equation}
where $B_i=B(\omega_i,x),$ and we abbreviate by $T_{jlmn}$ the kernel $T(k_j,k_l,k_m,k_n)$ of the Zakharov equation. In \eqref{eq:Spatial ZE Discr} the function $\delta$ is the ordinary Kronecker delta function. To extract the effect of nonlinear dispersion we follow the procedure outlined by Stuhlmeier \& Stiassnie \cite{Stuhlmeier2019} for the conventional temporal Zakharov equation (see e.g.\ Chapter 14 of Mei et al \cite{Mei2005} for background).

We write the complex amplitude $B_j(x)$ as $|B_j|\exp(i \arg B_j),$ where both magnitude and argument may depend on $x.$
Separating into real and imaginary parts leads to:
\begin{align}
& c_{g,j} \frac{d |B_j|}{dx} = - \sum T_{jlmn} \delta_{jl}^{mn} |B_l||B_m||B_n| \sin(\theta_{jlmn}),\\
& -c_{g,j} |B_j| \frac{d \arg(B_j)}{dx} = \sum T_{jlmn} \delta_{jl}^{mn} |B_l||B_m||B_n| \cos(\theta_{jlmn}), 
\end{align}
with
\begin{align*}
&\theta_{jlmn} = \Delta x + \arg B_j + \arg B_l -\arg B_m -\arg B_n,\\
&\delta_{jl}^{mn} = \delta(\omega_j + \omega_l - \omega_m - \omega_n), \text{ and } \\
&\Delta = k_j + k_l - k_m - k_n.
 \end{align*}
Assuming that there is negligible evolution of the amplitudes, so that the $|B_j|$'s may be replaced by their initial values $|B_j(0)|,$ we rewrite
\begin{equation}
-c_{g,j} \frac{d}{dx} (\arg B_j) = \frac{1}{|B_j|} \left( \sum_l e_{lj} |B_l|^2|B_j| T_{jljl}  + \sum_l \sum_{m \neq j} \sum_{n\neq j} T_{jlmn} \delta_{jl}^{mn} |B_l||B_m||B_n| \cos(\theta_{jlmn}) \right)
\end{equation}
and, neglecting the second term on the right-hand side (which captures only exactly resonant quartets) integrate:
\begin{equation}
-c_{g,j} (\arg B_j) = \sum_l e_{lj} |B_l|^2 T_{jljl}  x + \arg B_j(0),
\end{equation}
where $e_{np} = 1$ for $n = p$ and $e_{np} = 2$ for $n \neq p.$  The kernels of the Zakharov equation reduce for two unidirectional waves to 
\begin{equation} \label{eq: Zakharov kernel}
T(k,k_1,k,k_1) = \begin{cases} 
\frac{k_x k_{1x}^2}{4 \pi^2} & \text{ for } k_{1x} < k_x ,\\
\frac{k_x^2 k_{1x}}{4 \pi^2} & \text{ for } k_{1x} \geq k_x.
\end{cases}
\end{equation}

This leads to a correction for the wavenumber:
\begin{equation} \label{eq: wavenumber correction}
K_n = k_n - \frac{1}{c_{g,n}} \sum_l e_{ln} |B_l(0)|^2 T_{lnln}, \end{equation}
which is the counterpart to the well-known Stokes' correction to the frequency. The effects of (weak) nonlinearity are thus to decrease the wavenumber by an amount of $O(\epsilon^2)$ compared to the linear theory. We will explore the effect of this wavenumber correction on two explicit solutions below, and see that it also impacts the predictable region discussed in section \ref{ssec: Linear forecasting} above.

\subsubsection{Explicit solutions to the spatial Zakharov equation}

The spatial Zakharov equation \eqref{eq:Spatial ZE} can be easily solved in two special cases: a single mode, or two modes. The former corresponds to the spatial evolution of the well-known Stokes' wave \cite{Stokes1847}, and the latter to the spatial evolution of the third-order two-wave system first considered by Longuet-Higgins \& Phillips \cite{Longuet-Higgins1962c}. Either of these cases trivially fulfil the resonance condition, since $\delta(\omega_a + \omega_a -\omega_a -\omega_a )=1$ and $\delta(\omega_a +\omega_b-\omega_a -\omega_b  )=1.$ Because the viewpoint of wavenumber correction (rather than frequency correction) is somewhat unusual in water waves, it is instructive to consider these solutions.

In case of a single wave $\omega_j,$ the spatial Zakharov equation (with $T_{jjjj}$ abbreviated by $T_j$) becomes
\begin{equation}
i c_{g,j} \frac{d B_j(x)}{dx} = T_{j} |B_j(x)|^2 B_j(x),
\end{equation}
which admits the constant amplitude solution
\begin{equation} \label{eq: SZE Single Mode Solution}
B_j(x)= A_j e^{-i A_j^2 T_{j} x/c_{g,j}}. \end{equation}

If two waves are present, say $\omega_a$ and $\omega_b,$ the spatial ZE becomes the coupled system
\begin{align}
i c_{g,a} \frac{d B_a}{dx} &= T_{a} |B_a|^2 B_a + 2 T_{ab} |B_b|^2 B_a,\\
i c_{g,b} \frac{d B_b}{dx} &= T_{b} |B_b|^2 B_b + 2 T_{ab} |B_a|^2 B_b,
\end{align}
where the symmetry of the kernel $T_{abab}=T_{abba}$ has been used to simplify the expressions and $T_{abab}$ has been abbreviated by $T_{ab}.$ Again, this system admits a solution with constant amplitudes $A_a$ and $A_b,$ 
\begin{align}
B_a(x) &= A_a \exp ( - i (T_a A_a^2 + 2 T_{ab} A_b^2)x/c_{g,a}),\\
B_b(x) &= A_b \exp ( - i (T_b A_b^2 + 2 T_{ab} A_a^2)x/c_{g,b}).
\end{align}

The relationship between the complex amplitudes and the leading order free surface elevation is given by 
\begin{equation}
\eta(x,t) = \frac{1}{2\pi} \int_{-\infty}^{\infty} \left( \frac{\omega}{2g} \right)^{1/2} \left[ B(x,\omega) \exp(i(k(\omega)x-\omega t)) + \text{c.c.} \right] d \omega.
\end{equation}
Here ``c.c.'' stands for the complex conjugate of the preceding expression. For a single mode $B(x,\omega) = B_0(x)\delta(\omega-\omega_0)$ we find, using \eqref{eq: SZE Single Mode Solution},

\begin{equation}\eta(x,t) = \frac{1}{\pi}  \left( \frac{\omega_0}{2g} \right)^{1/2} A_0 \cos(([k(\omega_0)-A_0^2 T_{0000} x/c_{g,0}]x-\omega_0 t)). \end{equation}
This is a simple sinusoidal wave with a wavenumber altered due to the effect of the weakly nonlinear dispersion relation. Normalising the constant amplitude via
\[  A_0 = \pi a_0 \left( \frac{2g}{\omega_0} \right)^{1/2},\]
this reduces to
\[\eta(x,t) = a_0 \cos (k_0[1 - a_0^2 k_0^2]x-\omega_0 t). \]
An exactly analogous procedure for two waves ($k_a < k_b$) yields the free surface
\[ \eta(x,t) = a_a \cos\left(k_a\left[1 - a_a^2 k_a^2 - 2 a_b^2 k_a^{3/2} k_b^{1/2}\right]x - \omega_a t\right) + a_b \cos\left(k_b\left[1 - a_b^2 k_b^2 - 2 a_a^2 k_a^{3/2}k_b^{1/2}\right]x - \omega_b t\right). \]
This shows clearly that the dispersion of one mode $\omega_a$ is influenced both by its own steepness $a_a$ and also by that of the second mode $\omega_b.$ 
It is easy to verify that these wavenumber corrections can be obtained from equation \eqref{eq: wavenumber correction}, making liberal use of \eqref{eq: Zakharov kernel} to simplify the kernels.

\subsubsection{Nonlinear forecasts with wavenumber correction}
\label{sssec: Nonl forecasts with k correction}
The wavenumber correction \eqref{eq: wavenumber correction} gives rise to a simple improved forecast
\begin{equation} \label{eq: zeta nonlinear}
\zeta_N(x,t) = \frac{Y_0}{N} + \frac{1}{N} \sum_{m=1}^{N/2-1} \left[  Y_m \exp(i(K_m x - \omega_m t)) + Y^*_m \exp(-i(K_m x -  \omega_m t)) \right],
\end{equation}
where $K_n$ denotes the corrected wavenumber. This is otherwise cosmetically identical to the linear forecast \eqref{eq: zeta linear}, a kinship which demonstrates the advantages of the formulation. Indeed, the only additional computational cost consists of calculating the $K_n$'s from the Fourier amplitudes via simple algebra. 

In addition, the change in wavenumber leads to a change in phase and group velocities, which become $c_{p,j} = \omega_j/K_j$ and $c_{g,j}=d \omega_j/ dK_j,$ respectively. The latter of these must be evaluated numerically in practice. The nonlinear corrected velocities are somewhat larger, effectively because of a smaller denominator ($K_j$ is smaller than the linear wavenumber $k_j,$ and this effect is more pronounced for shorter waves than for longer waves), which means that the spatial predictable zone increases in size slightly when nonlinear corrections are taken into account. The extent of this increase depends on both the frequencies and the amplitudes present in the measured sea.

\section{Deterministic forecasting of synthetic seas}
\label{sec: Synthetic DWF}

To test our two deterministic forecasting methods it is necessary to generate a wave field, measure it at a point, and compare the forecast with other measurements. This procedure can be undertaken either in a wave flume or computationally, and we devote this section to the latter. The advantage of synthetic seas lies in the ease of tuning inputs, and the ability to easily produce many realisations of different cases. To this end, we employ the open source HOS-NWT code \cite{Ducrozet2012} which implements the high-order spectral method (HOS).

\begin{figure}
\centering
\includegraphics[scale=1]{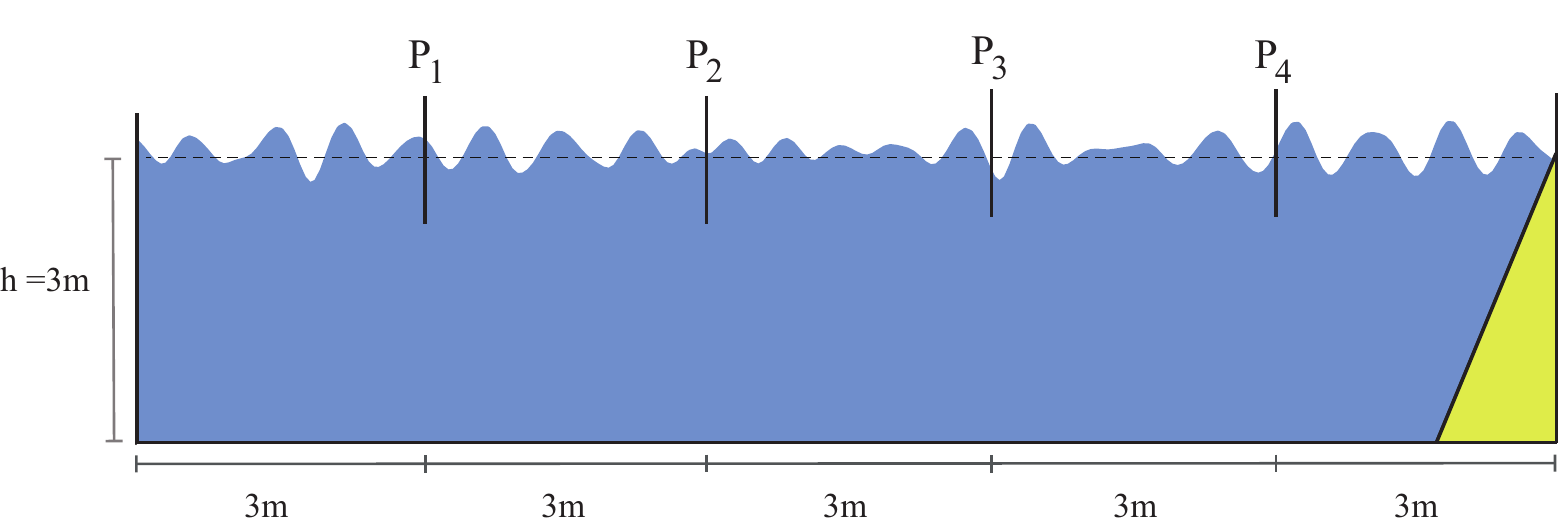}
\caption{Diagram of the HOS-NWT numerical wave flume configuration. $P_1$--$P_4$ denote the probes at which measurements are taken. Right-going waves are generated in water of 3 m depth, and fully absorbed by a numerical beach at the right-hand side of the flume.}
\label{fig: numerical flume}
\end{figure}

Throughout we use a numerical wave flume to generate purely unidirectional, random wave fields initialised by JONSWAP spectra with peak frequency $f_p = 1.3$ Hz, values of significant wave--height $H_s$ between 0.02 -- 0.04 m, and peak-sharpening parameters $\gamma=1,\, 3.3$ and 7. The numerical wave flume is 15 m long and 3 m deep, with probes located at 3, 6, 9 and 12 m, and has a fully absorbing beach, as depicted in figure \ref{fig: numerical flume}. HOS is used to model the nonlinear propagation of waves along the flume, and includes nonlinear dispersion, (near) resonant energy exchange, and the effects of bound modes -- considerably more physics than captured by either of our simple forecasts.

\begin{figure}
\centering
\includegraphics[width=0.5\linewidth]{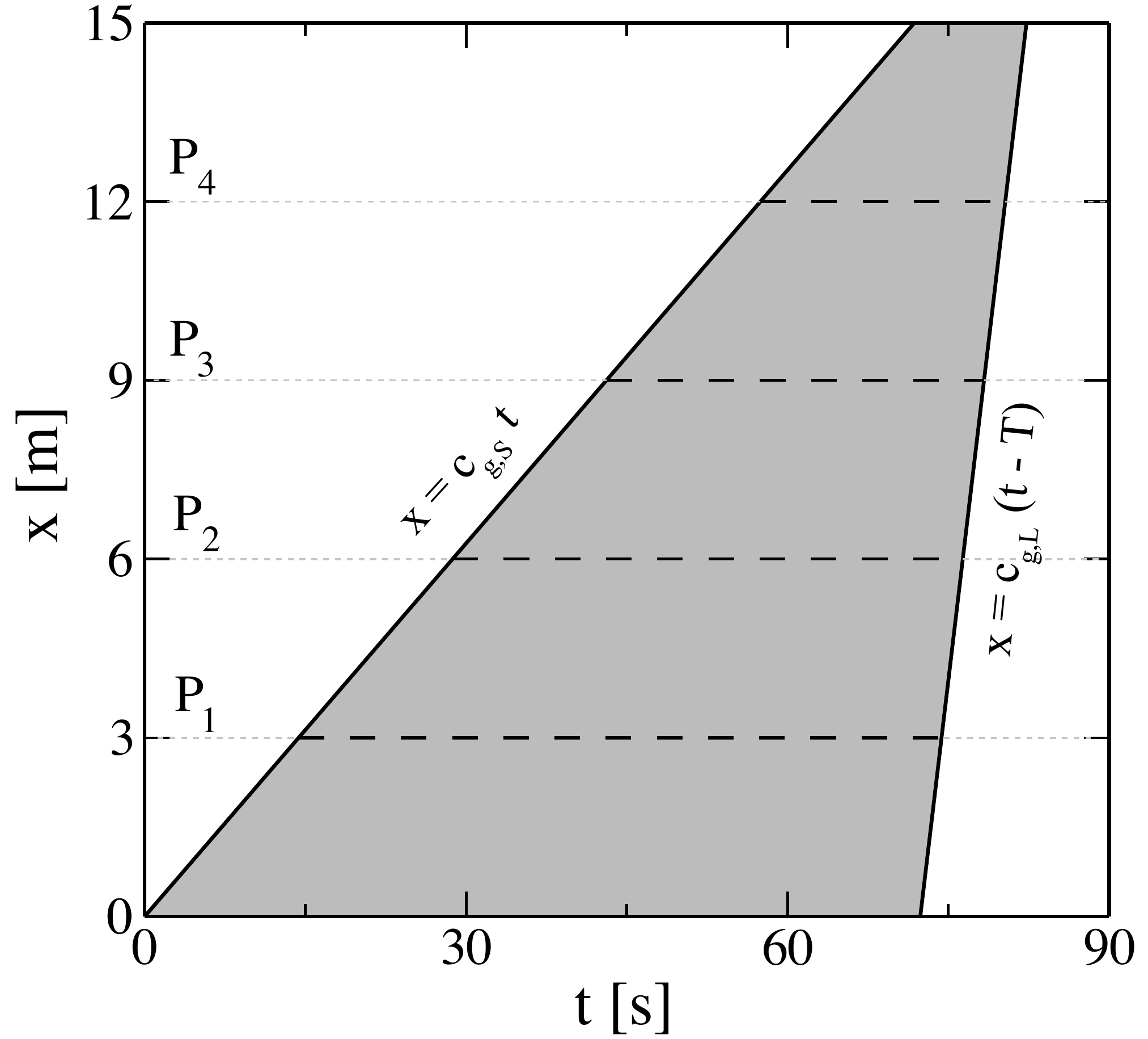}
\caption{The nonlinear predictable zone for a JONSWAP spectrum with $H_s = 0.03$ m, $\gamma=3.3$ generated in the 15 m long HOS-NWT numerical wave flume. Waves are generated at the wavemaker ($x=0$ m) between $t=0$ and $t=T=72$ s.}
\label{fig: example predictable zone}
\end{figure}

To produce a forecast, measurements from probe $P_1$ are sampled, the Fourier amplitudes extracted via FFT, and inserted into either \eqref{eq: zeta linear} or \eqref{eq: zeta nonlinear}. The resulting forecasts can then be compared to the measured time series at probes $P_2$, $P_3$, and $P_4$. Figure \ref{fig: example predictable zone} depicts the nonlinear predictable zone based on an example for $H_s = 0.03$ and $\gamma = 3.3.$ Waves are generated by the wavemaker at $x=0$ m and propagate along the 15 m long numerical flume. The longest waves resolved are $\omega_L=3.1 \, \text{rad}/s$ and the shortest $\omega_S=25.1 \, \text{rad}/s,$ and the associated nonlinear group velocities determine the edges of the predictable zone. The data at $P_1$ is then used to generate forecasts as shown in figure \ref{fig: forecast comparison Hs 0.03, gamma 3.3}.
The horizontal (time) axis has a different starting point for each panel of figure \ref{fig: forecast comparison Hs 0.03, gamma 3.3}, reflecting the narrowing of the predictable region seen in figure \ref{fig: example predictable zone} (for convenience we have here set $t=0$ s as the beginning of the record at $P_1,$ rather than the start of the wavemaker). 

\begin{figure}
\centering
\includegraphics[width=\linewidth]{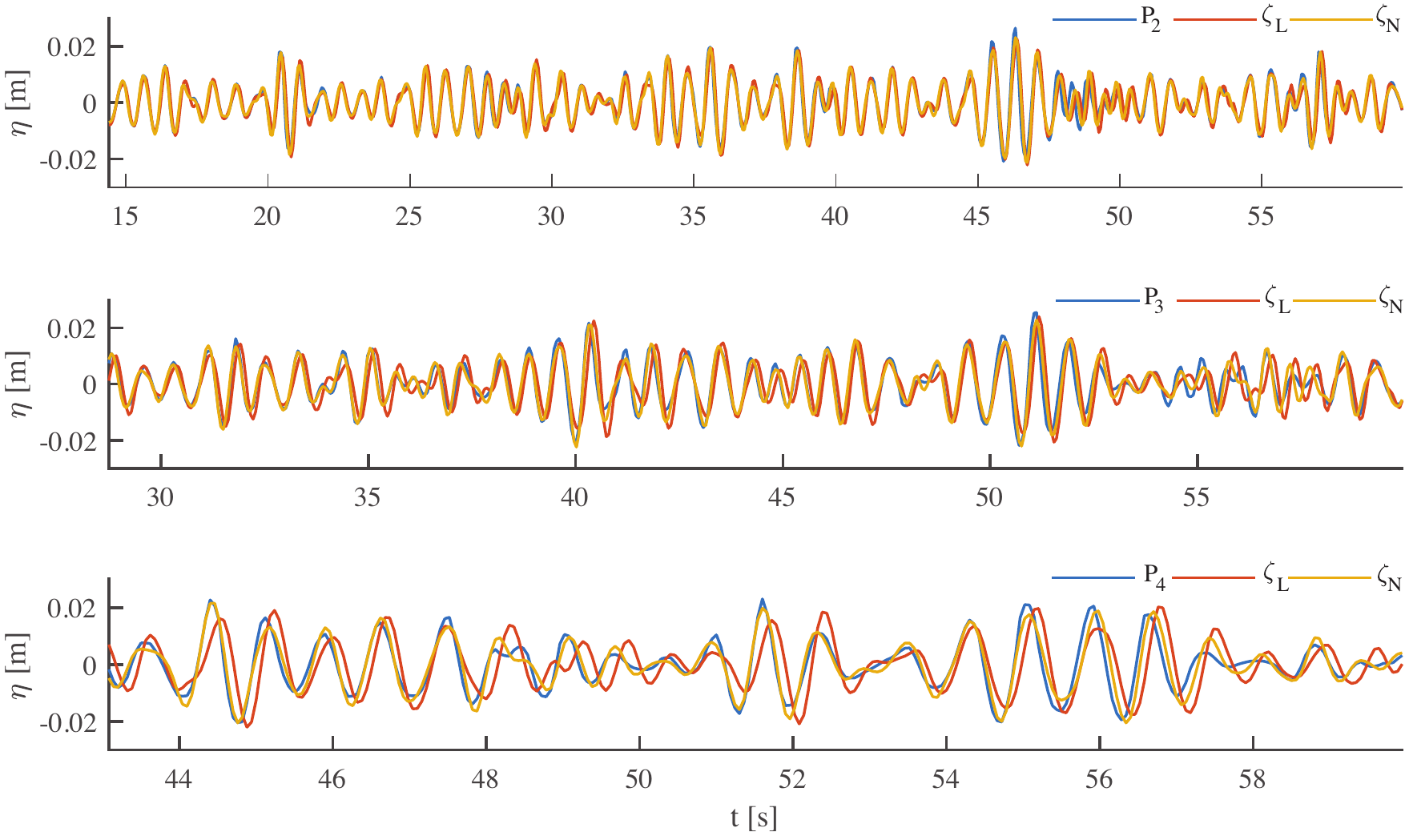}
\caption{Comparison of linear ($\zeta_L$, red curve) and nonlinear forecasts ($\zeta_N$, yellow curve) with measurements ($P_i,$ blue curve) for a numerically generated JONSWAP spectrum with $H_s=0.03$ and $\gamma=3.3.$ The horizontal axis shows time (a section through the nonlinearly corrected predictable region), while the vertical shows free surface elevation. (Top panel) comparison with measurements at probe $P_2.$ (Middle panel) comparison with measurements at probe $P_3.$ (Bottom panel) comparison with measurements at probe $P_4.$}
\label{fig: forecast comparison Hs 0.03, gamma 3.3}
\end{figure}

Figure \ref{fig: forecast comparison Hs 0.03, gamma 3.3} shows excellent agreement between measurement and both forecasts for the closest probe $P_2,$ while the forecast grows progressively less accurate as distance from the measurement point increases. However, the nonlinear forecast $\zeta_N$ remains considerably closer to the measured data at $P_4,$ some ten peak wavelengths from the initial measurement.

In order to obtain a measure of the aggregate quality of a forecast it is useful to compare the fit over several realisations. Table \ref{table:mean correlation HOS} considers the mean correlation (denoted $\bar{\rho}$) between a linear or nonlinear forecast produced from probe 1 and the predictable portion of the measured time series at probe $n,$ for $n=2,\, 3,$ or 4. Here and elsewhere the predictable region is calculated based on the nonlinear group velocities; it is therefore strictly slightly larger than the comparable predictable zone based on linear group velocities, see section \ref{sssec: Nonl forecasts with k correction}. In each case 20 realisations of a given JONSWAP spectrum with random, uniformly distributed phases are considered, and the forecasting procedure applied to each realisation in turn, before taking the arithmetic mean of the linear correlations.

\begin{table}[h]
\centering
\renewcommand\arraystretch{1.2}
\begin{tabular}{@{} l c c c  c c c c c c  @{}}
 & \multicolumn{3}{c}{$\gamma = 1$} & \multicolumn{3}{c}{$\gamma = 3.3$} & \multicolumn{3}{c}{$\gamma = 7$} \\
$H_s$ & 0.02 &0.03 & 0.04 & 0.02 & 0.03 & 0.04 & 0.02 & 0.03 & 0.04 \\
\cmidrule(lr){2-4}  \cmidrule(lr){5-7} \cmidrule(lr){8-10} 
$\bar{\rho}_{1,2}^L$ & 0.9663 & 0.8498 & 0.7204 & 0.9802 & 0.9015 & 0.8012 & 0.9823  & 0.9274 & 0.8382\\
$\bar{\rho}_{1,2}^{N}$ & 0.9883 & 0.9432 & 0.8914 & 0.9933 & 0.9637 &  0.9297 & 0.9928 & 0.9704 & 0.9368\\
\cmidrule(lr){2-4} \cmidrule(lr){5-7} \cmidrule(lr){8-10}
$\bar{\rho}_{1,3}^L$ & 0.8895 & 0.6375 & 0.4327 & 0.9373 & 0.7419 & 0.5323 & 0.9449 & 0.7930 & 0.6014\\
$\bar{\rho}_{1,3}^{N}$ & 0.9665 & 0.8736 & 0.7898 & 0.9838 & 0.9227 &  0.8391 & 0.9834 & 0.9311 & 0.8630\\
\cmidrule(lr){2-4} \cmidrule(lr){5-7} \cmidrule(lr){8-10}
$\bar{\rho}_{1,4}^L$ & 0.8096 & 0.4499 & 0.2574 & 0.8656 & 0.5604 &  0.3176 & 0.9036 & 0.6314 & 0.3667\\
$\bar{\rho}_{1,4}^{N}$ & 0.9530 & 0.8071 & 0.7057 & 0.9660 & 0.8708 &  0.7392 & 0.9762 & 0.8700 & 0.7693 \\
\bottomrule
\end{tabular}
\caption{Mean correlations $\bar{\rho}$ over 20 realisations between synthetic simulated seas and forecasts based on linear (superscript $L$) and nonlinear (superscript $N$) forecasts. Subscripts $i,j$ denote that measurements from probe $i$ are used to forecast probe $j.$ Significant wave height $H_s$ ranges from 0.02 to 0.04, and peak-sharpening parameters $\gamma=1, \, 3.3$ and 7.}
\label{table:mean correlation HOS}
\end{table}

As anticipated from figure \ref{fig: forecast comparison Hs 0.03, gamma 3.3}, the average correlation between measurement and forecast decreases with distance along the flume. For higher wave steepness, nonlinear dispersion is well captured by the corrected forecast $\zeta_N$ (see rows $\rho^N_{i,j}$), as evidenced by the good agreement for moderate distances. For the steepest waves and longest propagation distances the forecast quality degrades markedly, nevertheless the nonlinear forecast retains a clear advantage. Indeed, for $H_s=0.3$ and 0.04 and over all values of $\gamma,$ the nonlinear forecast can predict twice as far (6 m vs 3 m) as the linear forecast with the same average accuracy. It is also interesting to note that prediction is consistently easier for narrower spectra ($\gamma = 3.3, \, 7),$ with accuracy of both linear and nonlinear forecasts increasing at a given distance as $\gamma$ increases.

\section{Deterministic forecasting in a wave flume}
\label{sec: Experimental DWF}

\begin{figure}
\centering
\includegraphics[scale=1]{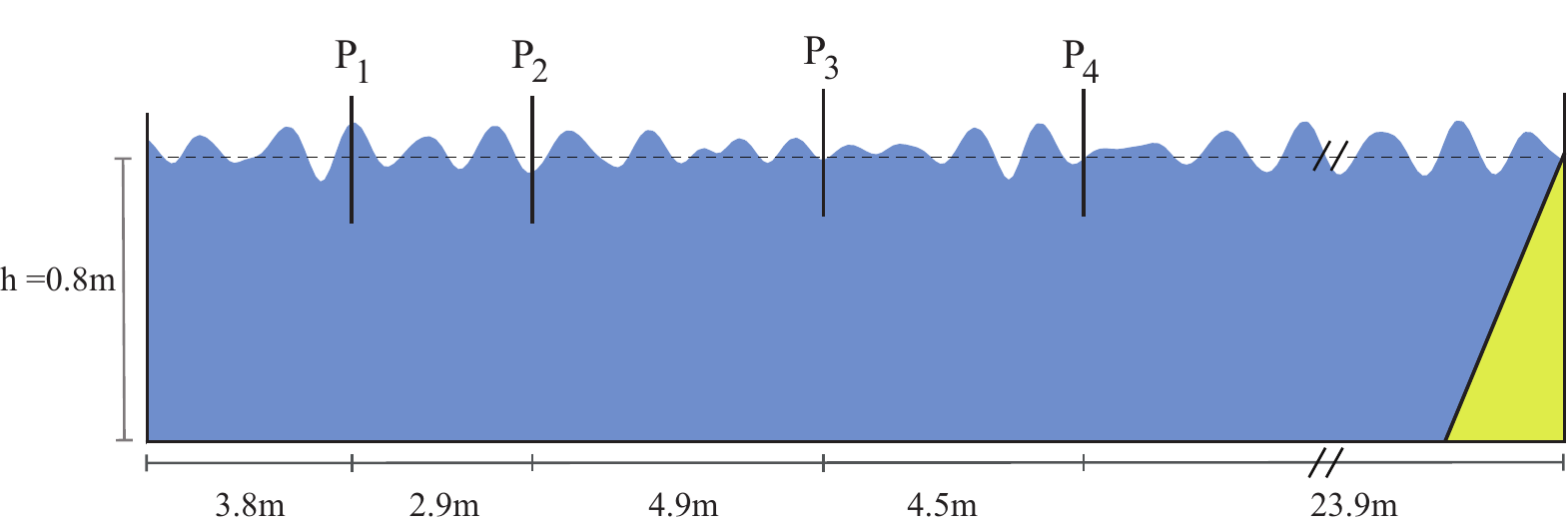}
\caption{Diagram of the experimental configuration. $P_1$--$P_4$ denote the probes at which measurements are taken. Right-going waves are generated in water of 0.8 m depth.}
\label{fig: experimental flume}
\end{figure}

To assess the accuracy of our new nonlinear forecasting approach, we also compare with experimental data from the 40 m long, 2.7 m wide flume at IRPHE/Pytheas Aix Marseille University. Data is taken from four probes placed at distances of 3.79, 6.64, 11.63, and  16.11 m from a piston wave maker in water of depth $d=0.8$ m, as shown in figure \ref{fig: experimental flume}. As above, measurements from probe 1 will supply the data necessary to produce a forecast, which is then compared with data from probes 2--4. We will consider three cases: (J1) a JONSWAP spectrum with $f_p = 1.10$ Hz, $\gamma = 3.3$ and $H_s = 0.01$ m, (J2) a steeper JONSWAP spectrum with $f_p = 1.11$ Hz, $\gamma= 3.3$ and $H_s =0.04$ m, and (M) a modulated plane wave, consisting of a plane wave with $f_p = 1.42$ and slope $ak = 0.16,$ and two side bands.

\begin{figure}
\begin{subfigure}[b]{\linewidth}
\centering
\includegraphics[width=\linewidth]{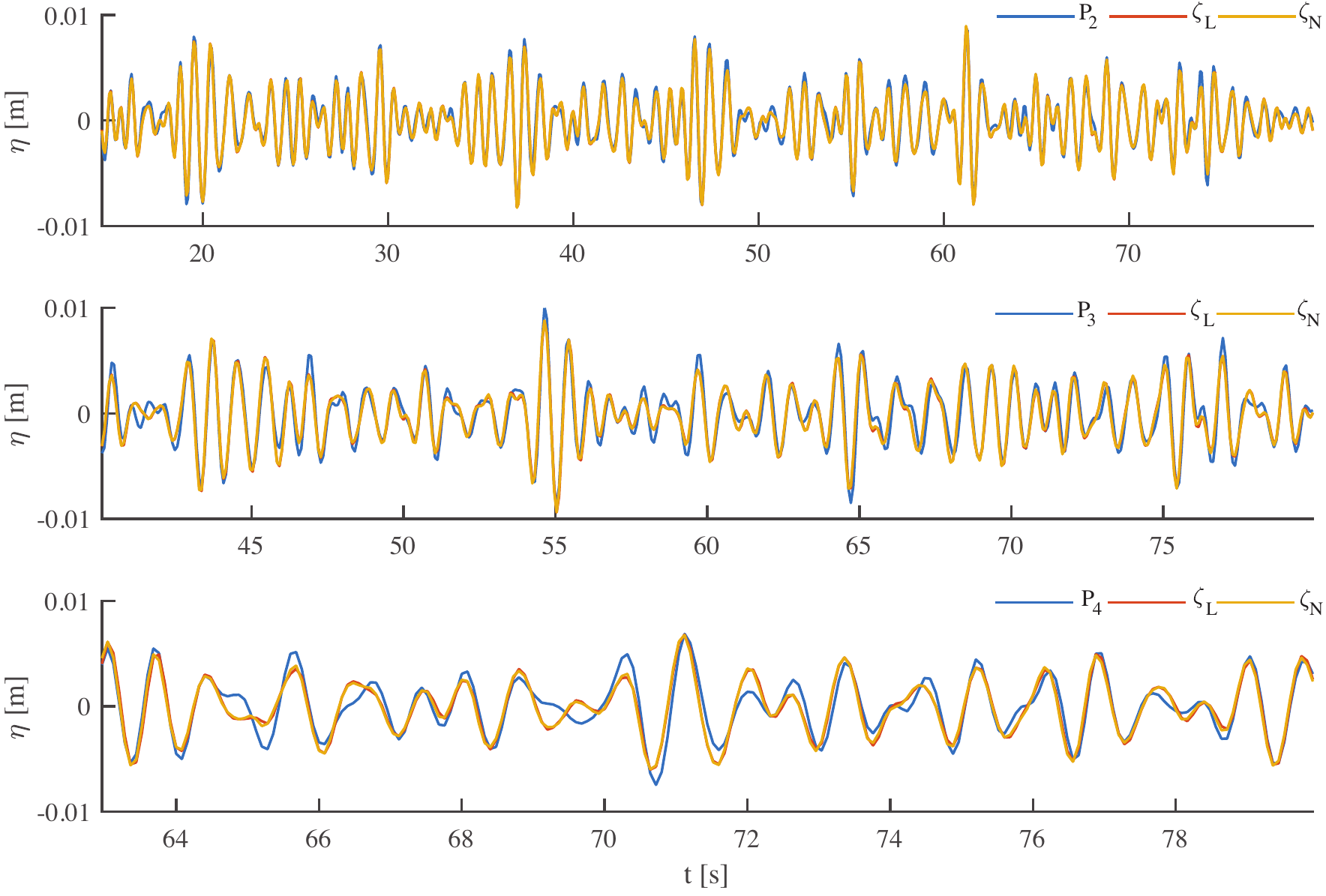}
\caption{JONSWAP Case J1.}
\label{fig: Comparison Case J1}
\end{subfigure}
\begin{subfigure}[b]{\linewidth}
\centering
\includegraphics[width=\linewidth]{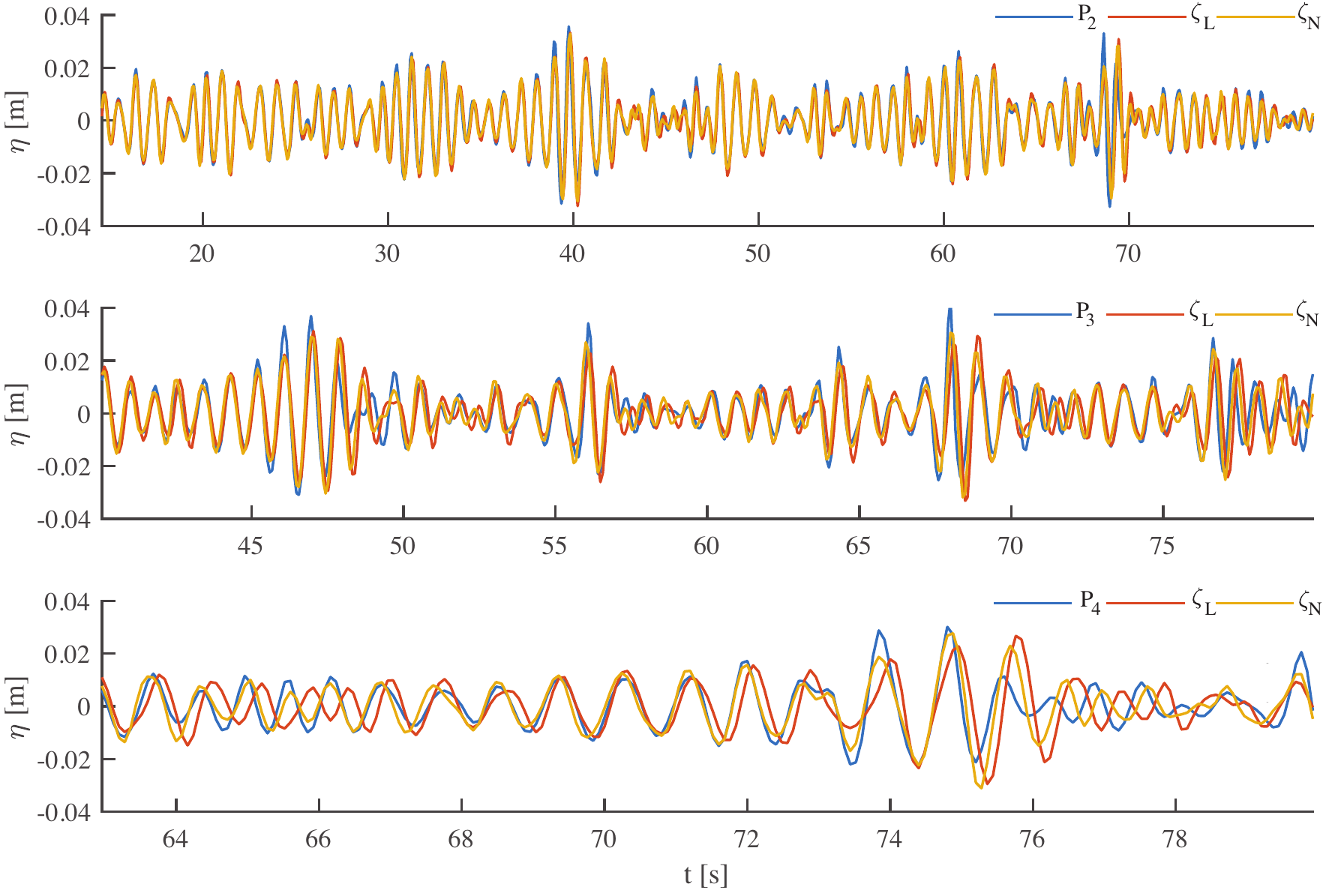}
\caption{JONSWAP Case J2.}
\label{fig: Comparison Case J2}
\end{subfigure}
\caption{Comparison of linear ($\zeta_L$, red curve) and nonlinear ($\zeta_N$, yellow curve) forecasts with measured probe data ($P_i,$ blue curve) for JONSWAP cases J1 and J2. Measurements taken at probe $P_1$ supply the Fourier amplitudes for the forecasts at probe $P_2$ (top panel), probe $P_3$ (middle panel) and probe $P_4$ (bottom panel).}
\label{fig: Comparison J1--J2}
\end{figure}

Forecasts for the two JONSWAP cases J1 and J2 are depicted in figures \ref{fig: Comparison Case J1} and \ref{fig: Comparison Case J2}. For mild seas with $H_s = 0.01$ either linear or nonlinear forecasting produces excellent agreement with measurements up to probe $P_3,$ nearly 8 m (ca.\ 7 peak wavelengths) away, as seen in figure \ref{fig: Comparison Case J1}. Due to the low steepness, the nonlinear correction is essentially negligible, and $\zeta_L$ is barely distinguishable from $\zeta_N.$ 
Akin to what was observed for synthetic data generated by HOS in section \ref{sec: Synthetic DWF}, as the wave steepness is increased to $H_s=0.04$ m the forecasts begin to depart from the measured data. The difference in linear and nonlinear forecasts is clearer here, with the quality of the dispersion-corrected forecast $\zeta_N$ outstripping the simple linear case $\zeta_L.$ This information is also captured by the correlation, used in table \ref{table: Experimental forecasting correlations} to provide a measure of forecast quality (here only over a single experimental realisation). 

\begin{figure}
\centering
\includegraphics[width=\linewidth]{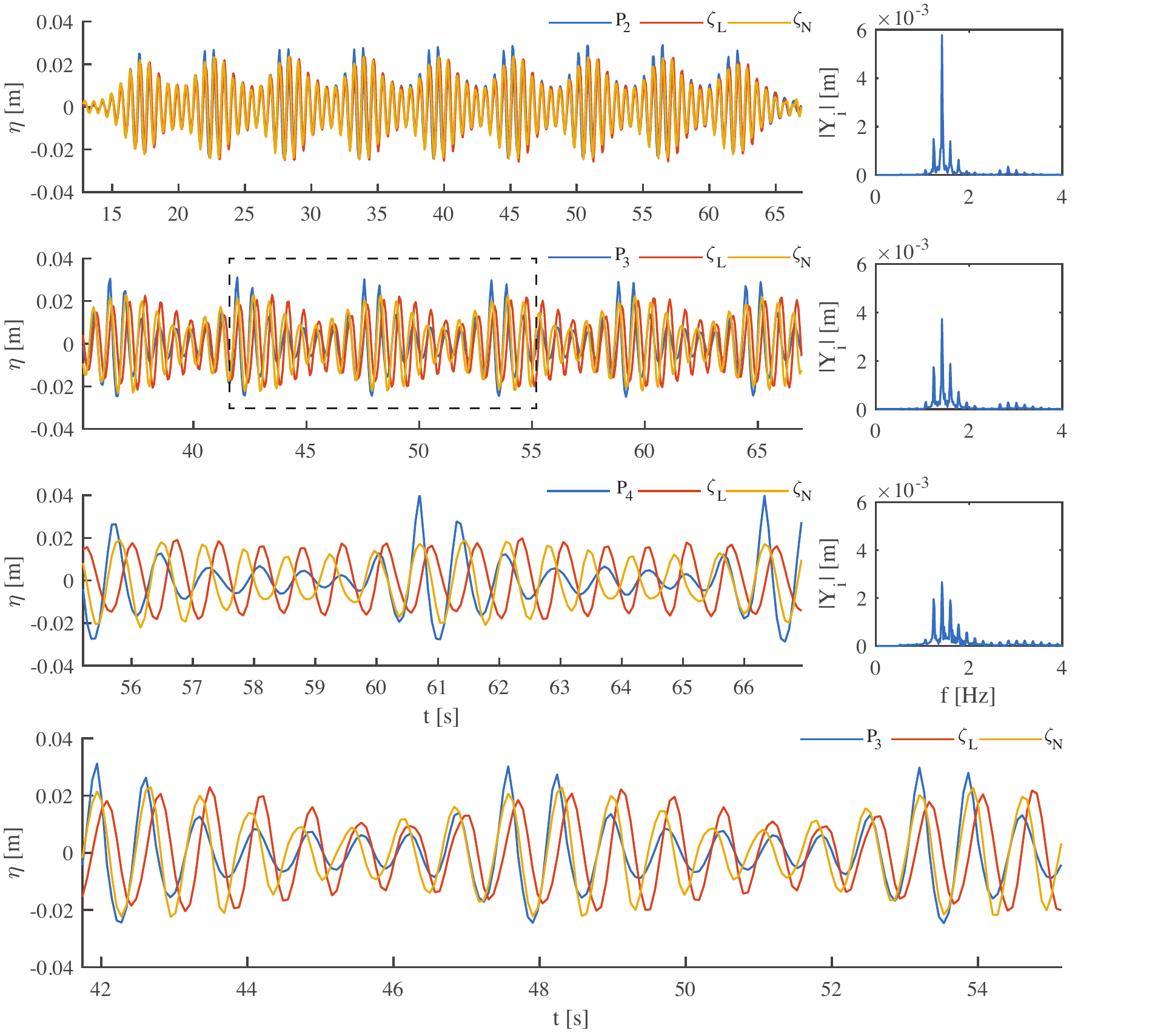}
\caption{Comparison of linear ($\zeta_L$, red curve) and nonlinear ($\zeta_N$, yellow curve) forecasts with measured probe data ($P_i,$ blue curve) for modulated plane wave case M. Measurements taken at probe $P_1$ supply the Fourier amplitudes for the forecasts at probe $P_2$ (top panel, 2.85 m propagation distance), probe $P_3$ (second panel, 7.84 m propagation distance) and probe $P_4$ (third panel, 12.32 m propagation distance). The bottom panel is an enlargement of the region outlined in black in panel 2. The Fourier amplitude spectrum at the three measurement gauges $P_2,\, P_3$ and $P_4$ is shown adjacent to panels 1--3.}
\label{fig: Comparison Case M}
\end{figure}

Forecasts for a modulated plane wave are shown in figure \ref{fig: Comparison Case M}. This case exhibits the well-known modulational instability of a degenerate quartet consisting of a carrier and two side bands, and the side-band growth with propagation distance can be clearly seen in the insets depicting the Fourier amplitude spectrum at probes $P_2, \, P_3$ and $P_4.$ As the wave-field propagates along the flume, the side-band amplitudes to grow at the expense of the carrier, while also influencing the modes' dispersion.

The nonlinear forecast $\zeta_N$ employs only the initial mode amplitudes to calculate the corrections to the dispersion relation (the procedure described in section \ref{sssec: spatial ZE} effectively neglects the energy exchange between modes, employing the Fourier amplitudes of $P_1$ throughout). The effect of this is clearly visible in the enlarged forecast in figure \ref{fig: Comparison Case M} (bottom panel) and the data in table \ref{table: Experimental forecasting correlations}: the nonlinear forecast $\zeta_N$ shows excellent agreement with the measured phases, but fails to capture the evolving amplitudes at larger distances. Because the correlation is insensitive to amplitudes (i.e.\ the signals $\sin(x)$ and $2 \sin(x)$ have correlation $\rho=1$), good agreement is found for all $\rho_{i,j}^N$ in table \ref{table: Experimental forecasting correlations}. For yet longer propagation distances, the coupling between the phase and amplitude evolution eventually degrades the otherwise good match between the phases in $\zeta_N$ and the experimental record.

\begin{table}[h]
\centering
\renewcommand\arraystretch{1.2}
\begin{tabular}{@{} l c c c  @{}}
& Case M & Case J1 & Case J2 \\
\cmidrule(lr){2-4}  
$\rho_{1,2}^L$ & 0.799 & 0.981 &  0.914  \\
$\rho_{1,2}^{N}$ & 0.967 & 0.982 &  0.955 \\
\cmidrule(lr){2-4} 
$\rho_{1,3}^L$ & 0.307 & 0.957 & 0.675 \\
$\rho_{1,3}^{N}$ & 0.837 & 0.950 &  0.871 \\
\cmidrule(lr){2-4} 
$\rho_{1,4}^L$ & -0.178 & 0.929 &  0.551 \\
$\rho_{1,4}^{N}$ & 0.663 & 0.925 & 0.854 \\
\bottomrule
\end{tabular}
\caption{Correlation between linear (superscript $L$) and nonlinear (superscript $N$) forecast for the three experimental cases M, J1, and J2. Subscripts $i,j$ denote that measurements from probe $i$ are used to forecast probe $j.$}
\label{table: Experimental forecasting correlations}
\end{table}

\section{Discussion}

We have derived a compact and theoretically simple wavenumber correction from the spatial Zakharov equation, and demonstrated its utility in simple cases of wave forecasting from synthetic and experimentally generated waves. We have seen that this method accurately captures the most important aspects of nonlinear dispersion, and is an spatial analogue of the techniques developed by Stuhlmeier \& Stiassnie \cite{Stuhlmeier2021} for the temporal forecasting problem. These corrections have applicability beyond the immediate context of deterministic forecasting, for example to the so-called Molin lensing effect \cite{Molin2005} which has recently been studied in the context of wave run-up \cite{Zhao2019}. 

Deterministic wave forecasting has numerous uses, ranging from the control of wave energy converters to improve power capture \cite{Li2012a,Huchet2021}, to ship motion forecasting for maritime operations \cite{Al-Ani2020, Kusters2019}. Wave prediction theories, such as that presented in this manuscript, are only one part of this story: ocean waves must first be properly measured, and considerable work is being undertaken on X-band marine radar \cite{Zinchenko2021,Previsic2021}. Measurement errors and sources of noise must be dealt with efficiently, as discussed recently by Desmars et al \cite{Desmars2020}, for example by continuous data assimilation and ensemble Kalman filtering \cite{Wang2020}. Subsequently prediction can be accomplished with a wide variety of propagation techniques.

We have used the fast Fourier transform throughout, and tacitly assumed that it introduces no appreciable errors into the forecasting methodology. This is not quite the case, as the sampled water surface is not a strictly periodic signal, and we have only a finite-length snapshot at each probe location. This induces a rectangular windowing and results in spectral leakage, recently addressed in the context of forecasting \cite{Hlophe2021, Abusedra2011}.  

It is interesting to observe that in our synthetic forecasts a decrease in spectral width (by increasing $\gamma$) increases the average accuracy of our forecasts, as measured by the linear correlation and presented in table \ref{table:mean correlation HOS}. Of the two phenomena associated with cubically nonlinear wave propagation in deep water, energy exchange is expected to be more significant for a narrow spectrum, while frequency correction is expected to be less significant. The former phenomenon is connected to the Benjamin-Feir index introduced by Janssen (see \cite{Janssen2004}), connecting spectral width and scale of nonlinearity to the appearance of modulational instability. The latter is a consequence of the asymmetry of \eqref{eq: wavenumber correction}: energy in long waves has a large effect on the dispersion of short waves, but not vice versa. For a broader spectrum, with energy distributed among modes further from the spectral peak (especially in higher frequencies), these dispersion corrections will therefore be more significant (see \cite{Stuhlmeier2019}).

The extremely narrow and discrete spectrum of the modulated plane wave in figure \ref{fig: Comparison Case M}, allows to track energy transfer clearly, and makes it a popular laboratory wave, although it is unlikely to be found on the ocean. The energy exchange associated with the modulational instability \cite{Benjamin1967a,Chabchoub2015} drives significant changes in the spectral amplitudes between one probe and the next, while both linear and nonlinear forecasts implicitly assume the spectrum remains unchanged throughout the propagation. Therefore, it is interesting to see that as expected, the amplitudes of the waves are not well predicted, yet, the phases are matched very well for the nonlinear prediction.

\section*{Acknowledgements}
RS gratefully acknowledges support by EPSRC grant EP/V012770/1 and a QJMAM Fund grant from the IMA. Research visits to the University of Plymouth by MG were supported by a QJMAM Fund grant from the IMA. DE acknowledges financial support from the Swiss National Science Foundation (Fellowship P2GEP2-191480). The authors thank Christopher Luneau and Hubert Branger from the IRPHE/Pytheas Aix Marseille University for their help during the experiments.

\end{document}